\documentclass[preprint]{aastex}
\usepackage{lineno}
\usepackage{xcolor}  

\begin{document}

\title {WASP 0346-21: An EL CVn-Type Eclipsing Binary with Multiperiodic Pulsations in a Triple System }
\author{Jae Woo Lee$^{1,2}$, Kyeongsoo Hong$^{1}$, Min-Ji Jeong$^{3}$, and Marek Wolf$^{2}$ }
\affil{$^1$Korea Astronomy and Space Science Institute, Daejeon 34055, Republic of Korea}
\affil{$^2$Astronomical Institute, Faculty of Mathematics and Physics, Charles University in Prague, 180 00 Praha 8, V Hole\v sovi\v ck\'ach 2, Czech Republic}
\affil{$^3$Department of Astronomy and Space Science, Chungbuk National University, Cheongju 28644, Republic of Korea}
\email{jwlee@kasi.re.kr}

\begin{abstract} 
VLT/UVES spectroscopic and TESS photometric observations for WASP 0346-21 allow the direct determination of its physical properties, 
along with the detection of a circumbinary object and oscillating signals. The high-resolution spectra yielded the radial velocities of 
all three stars and the atmospheric parameters of $T_{\rm eff,A}$ = 7225$\pm42$ K, [M/H] = 0.30$\pm$0.03 dex, and $v_{\rm A}$$\sin i$ 
= 78$\pm$5 km s$^{-1}$ of the primary component. The combined analysis of these observations resulted in the fundamental parameters 
of the eclipsing components and the third light of $l_3$ = 0.043$\pm$0.004, which is consistent with the light contribution of 
the tertiary star observed in the echelle spectra. WASP 0346-21 A resides within the overlapping main-sequence domain of $\delta$ Sct 
and $\gamma$ Dor variables, while the secondary component of $M_{\rm B}$ = 0.185$\pm$0.013 M$_\odot$, $R_{\rm B}$ = 0.308$\pm$0.023 R$_\odot$, 
$T_{\rm eff,B}$ = 10,655$\pm$146 K, and $L_{\rm B}$ = 1.09$\pm$0.17 L$_\odot$ matches well with the low-mass white dwarf (WD) model for 
$Z$ = 0.01, corresponding to the thick-disk population classified by the Galactic kinematics. Multifrequency analyses were performed on 
the residual TESS data after removing the binarity effects. The low frequencies around 26.348 day$^{-1}$ and 17.683 day$^{-1}$ 
are $\delta$ Sct pulsations originating from WASP 0346-21 A, and the high frequencies of 97.996 day$^{-1}$ and 90.460 day$^{-1}$ are 
considered to be extremely low-mass WD oscillations. These results demonstrate that WASP 0346-21 is a hierarchical triple system, consisting 
of an EL CVn binary with multiperiodic pulsations in each component and a distant outer tertiary. 
\end{abstract}

\section{INTRODUCTION}

White dwarfs (WDs) are the final remnants of stars with initial masses up to about 10 M$_\odot$, corresponding to 95$-$97 \% of all stars 
in our Galaxy. Many of them are clustered around $\sim$0.6 M$_\odot$ and harbor CO cores and H-rich envelopes. Considering the age of 
the Universe, the WD minimum mass that can be formed from the evolution of a single star is between 0.3 M$_\odot$ and 0.45 M$_\odot$ 
(Kilic et al. 2007). However, the mass distribution of the stellar remnants indicates the presence of a non-negligible portion of 
extremely low-mass (ELM) WDs with less than 0.3 M$_\odot$ (Kleinman et al. 2013; Kepler et al. 2017), which is insufficient mass to burn 
their He cores. The ELM WDs are thought to be possible products of binary star evolution through stable or unstable mass transfer 
(Istrate et al. 2016; Calcaferro et al. 2018; Li et al. 2019). Observationally, their companions are A/F main-sequence (MS) stars, 
millisecond pulsars, and WDs.

EL CVn-type stars are eclipsing binaries (EBs) composed of an ELM WD precursor (pre-He WD) and an A/F dwarf (Maxted et al. 2014). 
Their MS binary precursors are mainly predicted to have orbital periods shorter than $\sim$3 days (Lagos et al. 2020). The pre-He WDs 
spend their lives in a constant luminosity phase, moving almost consistently toward higher temperatures through stable H-shell burning 
prior to reaching the WD cooling track (Chen et al. 2017). The light curves of the EL CVn binaries display total eclipses at the primary 
minima due to the occultation of the pre-He WDs and ellipsoidal variations outside eclipses through the tidal deformation of the MS stars. 
Based on the light curve morphology, more than 80 EL CVn-like EBs have been discovered through various ground- and 
space-based photometric surveys (van Kerkwijk et al. 2010, Carter et al. 2011, Breton et al 2012, Maxted et al. 2014, Faigler et al. 2015, 
Rappaport et al. 2015, van Roestel 2017, Wang et al. 2019, Lee et al. 2020, Peng et al. 2024). Their orbital periods ($P_{\rm orb}$) 
range from 0.46 to 23.9 days, but most of them have $P_{\rm orb} \la 3$ days. Tokovinin et al. (2006) showed that 96 \% of 
the MS binary stars with $P_{\rm orb} < 3$ days are at least triplets. Lagos et al. (2020) found tertiary objects in all of 
their EL CVn targets in a range of $P_{\rm orb}$ = 0.793--2.173 days, which implies that short-period EL CVn EBs are possible triple systems. 

The number of EL CVn candidates is steadily increasing, but with the exception of six double-lined EBs, most have no or insufficient 
spectroscopic data to reliably measure their physical properties. Some of them are known to display different classes of pulsations: 
$\delta$ Sct and/or $\gamma$ Dor variables in the A/F primary components, and very short-period oscillations of a few hundred seconds in 
the pre-He WD companions (cf. Hong et al. 2021; Kim et al. 2021). Such pulsating EL CVn EBs are promising targets for detailed studies of 
ELM WDs and intermediate-mass MS stars because of the strong synergy of their asteroseismology and binary properties. This article focuses 
on WASP 0346-21 (1SWASP J034623.68-215819.5; TIC 121078334, HD 23692, TYC 5887-800-1, Gaia DR3 5087757377681887232; $T_{\rm p}$ = $+$9.343; 
$V\rm_T$ = $+$9.620, $(B-V)\rm_T$ = $+$0.337). We found significant pulsation signals in the time-series TESS data of this target. 

Using archival WASP data analysis, Maxted et al. (2014) announced WASP 0346-21 to be a potential EL CVn-type EB with the following 
light curve parameters: an orbital period of $P_{\rm orb}$ = 0.9285752 days, an inclination of $i$ = 79.1 deg, a mass ratio of 
$q_{\rm ph}$ = 0.165, a light ratio of $L_{\rm B}/L_{\rm A}$ = 0.08, and relative radii of $r_{\rm A}$ = 0.3650 and $r_{\rm B}$ = 0.0628. 
They also estimated the effective temperatures of the binary components to be $T_{\rm eff,A}$ = 7400$\pm200$ K and $T_{\rm eff,B}$ = 
9950$\pm400$ K from a model fit to the observed flux distribution. We aim to explore the pulsation characteristics of WASP 0346-21 and 
to determine its stellar properties from a combined modeling of the TESS photometric data and the archival echelle spectra. 
These provide valuable information for understanding how short-period binaries with ELM WDs form and evolve, and for probing their interiors 
using asteroseismic tools. Throughout the paper, the primary and secondary components represent the more massive MS star (A) and 
hotter WD companion (B), respectively.

\section{TESS PHOTOMETRY AND ECLIPSE TIMINGS}

WASP 0346-21 was observed during Sectors 4 (S4) and 31 (S31) of the TESS mission (Ricker et al. 2015). The S4 observations were made in 
a 2-min cadence mode from October 19 to November 14 2018, while the S31 ones were obtained in both 20-s and 2-min modes from October 22 
to November 16 2020. The observational gap between S4 and S31 is approximately 708 days. This study utilized the \texttt{SAP$_{-}$FLUX} 
data collected from the MAST archive\footnote{\url{https://archive.stsci.edu/}. The TESS data used are available at \url{http://dx.doi.org/10.17909/ak4z-xn20}}. 
The raw measurements were detrended and normalized by applying a quadratic fit to the outside-eclipse portions of the TESS light curves. 
These data on the flux scale were converted to magnitude units by taking the average brightness at two quadratures as $T_{\rm p}$ = $+$9.343 
(Paegert et al. 2022). As a consequence, 31,441 and 95,045 individual points were secured in 2-min and 20-s observing modes, respectively, 
with a total period of nearly 760 days. The CROWDSAP values\footnote{fraction of target flux measured in the photometric aperture.} for 
S4 and S31 are 0.99725157 and 0.99575806, respectively, indicating that the TESS measurements of WASP 0346-21 have little contamination 
by nearby sources. 

The resultant time-series data of WASP 0346-21 are plotted in Figure 1, where the top panel shows 2-min cadence data from both sectors 
and the other panels detail 20-s cadence measurements from S31. In this figure, our target star appears as a detached binary exhibiting 
different eclipse depths and short periodic oscillations. Using the 2-min and 20-s cadence data, respectively, from S4 and S31, we fitted 
each eclipse curve with the method of Kwee \& van Woerden (1956), and obtained 47 primary and 48 secondary mid-eclipse times, compiled in 
Table 1. The primary minima were used to yield the linear ephemeris suitable for the TESS light curve of WASP 0346-21, as follows: 
\begin{equation}
\mbox{Min I} = \mbox{BJD}~ 2459145.207499(\pm0.000083) + 0.92857541(\pm0.00000035)E.  
\end{equation}
The TESS orbital period agrees within error with that (0.9285752$\pm$0.0000003 days) of Maxted et al. (2014) found in archival WASP data. 
However, when the TESS data are phased with the epoch of HJD 2455178.3330$\pm$0.0002 in Table 2 of Maxted et al. (2014), their orbital 
phases are shifted by about 0.18. This may be the result of changes in the orbital period of WASP 0346-21. Currently we do not have enough 
minimum epochs to check this possibility. Future continuous and systematic mid-eclipse timings are needed to study the orbital period change.

\section{VLT/UVES SPECTROSCOPY AND DATA ANALYSIS}

Spectroscopic observations for WASP 0346-21 were conducted from 2014 October 2 to 9 with the VLT 8.2-m telescope and the high-resolution 
UVES spectrograph (Dekker et al. 2000) at Paranal Observatory in Chile. The echelle spectra were made sequentially using the UVES blue 
and red arms in the wavelength ranges of 3282--4563 $\rm \AA$ and 5655--9464 $\rm \AA$, with resolving powers $R$ of $\sim 49,620$ and 
$\sim 51,690$, respectively. We collected the spectra processed by the UVES pipeline from the ESO portal\footnote{\url{https://archive.eso.org/scienceportal/home}}. 
The reduced spectra were formed by co-adding and merging three 120-s exposures. Most of them, with an integration time of 360 s, have 
a signal-to-noise ratio (SNR) of around 50--70. The normalization of the archival data was performed by fitting a cubic spline. 
The Barycentric Julian Date (BJD) and velocity corrections were applied to the UVES spectra. 

In EL CVn-type binaries, the spectral features of the secondary components are difficult to measure due to the low luminosity of the stars. 
Phase-folded trailed spectra are useful for tracking and revealing absorption lines caused by the orbital motions of binary components 
(e.g., Hong et al. 2015; Lee et al. 2020). Therefore, we closely examined the trailed spectra to find sets of absorption lines where 
both components are observable. The result showed that the absorption lines of the component stars clearly exist in two specific regions 
of 4440--4520 $\rm \AA$ (blue; \ion{Mg}{2} 4481 $\rm \AA$) and 6310--6390 $\rm \AA$ (red; \ion{Si}{2} at 6347 $\rm \AA$ and 6371 $\rm \AA$).

To extract the radial velocities (RVs) of WASP 0346-21, we employed the \texttt{RaveSpan} code (Pilecki et al. 2017). This program 
provides three methods for RV measurements: broadening function (BF; Rucinski 2002), simple cross-correlation (CCF; Simkin 1974; 
Tonry \& Davis 1979), and 2D cross-correlation (TODCOR; Mazeh \& Zucker 1994). In this study, we applied the BF method to 
all archival spectra, which showed superior results compared to CCF and TODCOR in preliminary RV measurements. The template spectrum 
for this technique was adopted from the library of synthetic local thermodynamic equilibrium (LTE) spectra by Coelho et al. (2005). 
The BF profiles for three orbital phases are illustrated as samples in Figure 2. Except for the BFs obtained at the conjunction phase 
of 0.013, the others exhibit three apparent peaks. Unlike the primary and secondary peaks, which show changes in RVs, 
the central third peaks appear to be constant throughout the observing runs. This implies that the additional sharp line profiles 
originate from a tertiary object and that WASP 0346-21 is a triple-lined RV system (cf. Lu et al. 2001). 
We measured the weighted mean RVs in each spectral region by fitting Gaussian and rotational profile functions to the peaks in 
the BFs several times. These are presented in Table 2 and depicted in Figure 3. We applied a sine wave to each RV curve of 
the binary components, and obtained a spectroscopic mass ratio of $q_{\rm sp}$ = 0.1179 from the sinusoidal semi-amplitudes of 
$K_{\rm A}$ = 27.6 km s$^{-1}$ and $K_{\rm B}$ = 234.1 km s$^{-1}$. 

We calculated the atmosphere parameters of WASP 0346-21 A by comparing synthetic spectral grids formed from all possible combinations with 
the UVES spectrum observed at phase 0.013, where the secondary companion is completely hidden behind the primary star. For this work, 
we used the Grid Search in Stellar Parameters (GSSP\footnote{The GSSP package is available at \url{https://fys.kuleuven.be/ster/meetings/binary-2015/gssp-software-package}}) 
software package by Tkachenko (2015). This package program allows five key stellar parameters to be optimized at once: effective temperature 
($T_{\rm eff}$), surface gravity ($\log g$), metallicity ([M/H]), projected rotational rate ($v$$\sin i$), and microturbulence velocity 
($v_{\rm{mic}}$). The synthetic spectra were generated by applying stellar atmosphere models computed with the \textsc{LLmodels} code 
(Shulyak et al. 2004) to the radiative transfer code \textsc{SynthV} (Tsymbal 1996), both of which are in the GSSP package. 
Maxted et al. (2014) used models fitted to the observed flux distribution and inferred the primary's temperature $T_{\rm eff,A}$ to be 
$7400\pm200$ K. Thus, for the synthetic stellar models the temperature range was given as 6800--8000 K with a step of 100 K. 
The surface gravity was assigned $\log$ $g_{\rm A}$ (cgs) = 4.13, calculated from our binary modeling in the next section. The micro and 
macro turbulence velocities were kept fixed at 1.0 km s$^{-1}$ and 4.21 km s$^{-1}$, respectively, by the empirical relations provided in 
the \texttt{iSpec} code (see Blanco-Cuaresma et al. 2014). [M/H] and $v_{\rm A}$$\sin i$ ranges were applied from $-0.8$ to $+0.8$ dex in 
0.1 dex steps and 60--100 km s$^{-1}$ in 1 km s$^{-1}$ steps, respectively. Initially, the UVES spectrum was automatically normalized 
using the SUPPNet software\footnote{The SUPPNet is available at \url{https://rozanskit.com/suppnet/}} (R{\'o}{\.z}a{\'n}ski et al. 2022). 

The best fitting model was obtained by applying the $\chi^2$ minimization method to the observed spectrum and the grids of synthetic spectra. 
In this calculation, we used the UVES blue part of the spectral range 3700--4500 $\rm \AA$, which contains unique absorption lines (H$_9$, 
H$_8$, \ion{Ca}{2} K, H$_{\epsilon}$, H$_{\rm \delta}$, H$_{\rm \gamma}$) for determining atmospheric parameters. Then, the observed spectrum 
was manually re-normalized by comparison with the resulting synthetic spectrum using the SUPPNet software. This grid search process was 
repeated until the $\chi^2$ value among the GSSP results was minimal. The optimal values of $T_{\rm eff,A}$, [M/H], and $v_{\rm A}$$\sin i$ 
were determined by polynomial fitting the lowest $\chi^2$ values in grids of each parameter obtained from the $\chi^2$ statistics, and 
their errors were estimated from the corresponding $\pm1\sigma$ confidence level values (cf. Lehmann et al. 2011): 
$T_{\rm eff,A}$ = 7225$\pm42$ K, [M/H] = 0.30$\pm$0.03 dex, and $v_{\rm A}$$\sin i$ = 78$\pm$5 km s$^{-1}$. Figure 4 shows the UVES spectrum 
at the primary eclipse and its optimal model.

\section{BINARY MODELING}

High-precision TESS data for WASP 0346-21 showed nearly identical light curves for both S4 and S31. The space-based observations present 
an ellipsoidal variation outside eclipses and a box shape with overlapping relatively very short oscillations in the primary eclipses. 
The light curve shape resembles that of the EL CVn type (Maxted et al. 2014; Wang et al. 2020). Maxted et al. (2014) analyzed 
the WASP survey data and the observed flux distribution of WASP 0346-21, and classified the variable star as an EL CVn-type binary with 
typical characteristics of a short $P_{\rm orb}$ = 0.9285752 days and a low $q_{\rm ph}$ = 0.165. However, the spectroscopic elements 
are not known for this system, so its binary properties are considered preliminary. To reliably determine a binary star model for 
WASP 0346-21, we analyzed in detail the archival TESS data and spectroscopic measurements using the Wilson-Devinney (W-D) program 
(Wilson \& Devinney 1971; Kallrath 2022) that is based on the Roche model and allows for proximity effects in the RVs. 

The combined synthetic modeling of the two datasets was performed in an almost identical way for WASP 0131+28 (Lee et al. 2020). 
In this synthesis, the surface temperature and rotation parameter of the cooler but more massive component were set to be $T_{\rm eff,A}$ 
= 7225 $\pm$ 42 K and $F_{\rm A}$ = 0.819$\pm$0.071, respectively, obtained from synthetic model fits to the observed UVES spectrum. 
The secondary companion WASP 0346-21 B was assumed to be in synchronous rotation ($F_{\rm B}$ = 1.0) after some trials. The logarithmic law 
for limb darkening was applied and its coefficients $x$ and $y$ were adopted from the updated values of van Hamme (1993). 
The gravity-darkening $g_{\rm A,B}$ and albedo $A_{\rm A,B}$ parameters were all assigned as 1.0 from the components' temperatures. 

In general, a simultaneous binary modeling of light and RV curves is possible. However, as previously mentioned, the orbital period 
of WASP 0346-21 may have changed, and there is a time gap of about 1471 days between the RVs and TESS measurements. For this reason, 
our combined analysis was carried out through two steps (Lee et al. 2018): in the first step, the time-series TESS data were analyzed 
applying the spectroscopic results. In the second, the ground-based RV curves were analyzed with the light curve solution computed 
in the first step. The modeling process was calculated iteratively until the corrections of the free parameters were smaller than 
their corresponding uncertainties and both types of observations were reasonably fitted. The binary modeling was performed for 
various Roche models, but only detached configurations showed satisfactory results. The best-fit solutions from this modeling are 
summarized in Table 3. The synthetic RV and light curves are illustrated as red solid curves in Figures 3 and 5, respectively. 
The third light of $l_3\approx$ 4.3 \% was detected in the binary modeling, which may result from the presence of a tertiary object, 
as discussed in the last section.

The combination of the light and RV parameters yielded the absolute dimension for WASP 0346-21 collected in Table 4. This calculation 
used $T_{\rm eff}$$_\odot$ = 5780 K and $M_{\rm bol}$$_\odot$ = +4.73 as the solar surface temperature and bolometric magnitude, 
respectively. We took the empirical bolometric corrections (BCs) expressed as a function of temperature $\log T_{\rm eff}$ (Flower 1996; 
Torres 2010). The physical properties of WASP 0346-21 A and B correspond approximately to an F0 main sequence and an ELM WD, respectively. 
The distance to the WASP 0346-21 system was measured by adopting $V$ = 9.581 $\pm$ 0.03 and $E$($B-V$) = 0.033 $\pm$ 0.006 from the TESS 
Input Catalogue (Paegert et al. 2022), and by applying $A_{\rm V}$ $\simeq$ 3.1$E(B-V)$ and $l_3$ = 0.0425. Our EB-based distance of 
269 $\pm$ 8 pc agrees within a 2-$\sigma$ error with 290 $\pm$ 5 pc taken from the Gaia DR3 parallax of 3.447 $\pm$ 0.063 mas (Gaia Collaboration 2022).

\section{PULSATIONAL CHARACTERISTICS}

The more massive primary star of WASP 0346-21 lies inside the main-sequence instability domain of A/F-type pulsating variables on 
the Hertzsprung-Russell (H-R) diagram, and the hotter secondary companion is located among extremely low-mass pre-He WD variables 
(pre-ELMVs; C\'orsico et al. 2019; Hong et al. 2021, Lee et al. 2022a). For a reliable frequency analysis, we subtracted 
our binary model in Table 3 from the time-series TESS observations, and analyzed the light curve residuals with iterative, 
sequential pre-whitening on the amplitude spectrum using the PERIOD04 software (Lenz \& Breger 2005). Details of this procedure were 
presented in a paper by Lee et al. (2014). Figure 6 displays the corresponding residuals for the full 2-min cadence data, clearly showing 
the multiperiod oscillations. 

To see if there are any changes in oscillation frequencies and amplitudes between S4 and S31, with a time gap of about 2 years, 
and between two exposure times of 2 min and 20 s, we divided the entire frequency range into two overlapping parts of 0$-$80 day$^{-1}$ 
and 70$-$$f_{\rm Ny}$, and separately analyzed the overall residuals by each sector and cadence. The Nyquist limit $f_{\rm Ny}$ is 
$\sim$360 day$^{-1}$ and $\sim$2164 day$^{-1}$, respectively, in 2-min and 20-s cadences. In this process, we extracted all peak signals 
with SNR $>$ 4 (Breger et al. 1993), and used 1.5 times the Rayleigh criterion of $\Delta f_{\rm res}$ = 1.5/$\Delta T$ = 0.058 day$^{-1}$ 
for frequency resolution (Loumos \& Deeming 1978), where $\Delta T$ = 26 days is the time span of each sector used. 
The amplitude spectra for three datasets (2-min S4 and S31, and 20-s S31) are displayed in the first to third panels of Figure 7. We can 
see strong low-frequency signals between 15 day$^{-1}$ and 30 day$^{-1}$, and noticeable signals, albeit with significantly lower amplitude, 
between 90 day$^{-1}$ and 100 day$^{-1}$. 

Ten significant frequencies with amplitudes higher than $\sim$0.4 mmag were detected identically in the three separate datasets and agreed 
with each other within a 0.01 day$^{-1}$ interval, which is about 6 times less than the frequency resolution of $\Delta f_{\rm res}$. 
In the 2-min cadence data, the amplitudes of the two most dominant frequencies, 26.348 day$^{-1}$ and 17.683 day$^{-1}$, 
were suppressed by approximately 0.66 mmag and 0.78 mmag, respectively, in S31 compared to S4, and the 97.996 day$^{-1}$ and 90.460 day$^{-1}$ 
signals in the high-frequency region were reduced by 0.005 mmag and 0.010 mmag, respectively, within their 1-$\sigma$ errors. 
On the other hand, the 2-min and 20-s exposure time data for Sector 31 had amplitudes that were consistent with each other within 
uncertainties. In sum, there were no significant differences in the frequencies and high-frequency amplitudes between datasets, 
but the dominant low frequencies differed significantly in amplitudes between S4 and S31. 

The bottom panel in Figure 7 is the PERIOD04 periodogram for all 2-min cadence residuals combined from both S4 and S31. We extracted 
signals with a SNR greater than 5 as possible pulsation frequencies, which is a detection threshold corresponding to a 0.1 \% false 
alarm probability for 2-min cadence TESS data simulated by Baran \& Koen (2021). A total of 58 candidate pulsations were acquired, and 
their frequencies $f_i$, amplitudes $A_i$ and phases $\phi$ are summarized in Table 5. The model curve synthesized from 
all extracted frequencies is shown as a red solid on the lower panel in Figure 6. Possible combination terms for these signals were 
carefully examined. We used the frequency resolution of $\Delta f_{\rm res}$ = 0.058 day$^{-1}$ corresponding to the duration of one sector, 
because the 2-year observational gap between S4 and S31 was too long (Breger \& Bischof 2002; Kim 2023). The resultant combinations are 
presented in the last column of Table 5. 

We believe that the four frequencies ($f_{11}$, $f_{16}$, $f_{41}$, $f_{47}$) below 4 day$^{-1}$ are likely to be orbital harmonics 
or instrumental artifacts that were not removed by detrending. Excluding these sidelobes, the low frequencies correspond to pulsation 
periods of $P_{\rm pul}$ = 37$-$100 min. Their pulsation constants and period ratios are in ranges of $Q$ = 0.014$-$0.038 days and 
$P_{\rm pul}/P_{\rm orb}$ = 0.028$-$0.075, respectively, which are typical values for $\delta$ Sct pulsators in EBs (Zhang et al. 2013; 
Kahraman Ali\c cavu\c s et al. 2017). The pulsation constants of the two strongest frequencies $f_{1}$ and $f_{2}$ are about 0.021 days 
and 0.031 days, respectively. Compared to 1.5 $M_\odot$ models of Fitch (1981), they can be identified as fundamental (F) and 
second harmonic (2H) radial $p$-modes, in the same order.

\section{DISCUSSION AND CONCLUSIONS}

We present a characterization of WASP 0346-21 by analyzing the VLT/UVES high-resolution spectra and the TESS photometric data in S4 
and S31. The BF profiles for the echelle spectra revealed the presence of a new tertiary object in addition to the two binary components. 
We measured the RVs of the three stars by the BF fits. The third body's RVs exhibit an average velocity of 36.7$\pm$0.6 km s$^{-1}$ 
during the 7.2-day spectroscopic observation period, with no trend in observation time or orbital phase. Using the GSSP package, 
we compared the observed UVES spectrum at the primary eclipse with all possible synthetic models, and utilized 
the $\chi^2$ minimization method to find the atmospheric parameters of $T_{\rm eff,A}$ = 7225$\pm42$ K, [M/H] = 0.30$\pm$0.03 dex, 
and $v_{\rm A}$$\sin i$ = 78$\pm$5 km s$^{-1}$. 

The light and RV curve analysis combined with the atmospheric parameters allows us to characterize each component of the EB system. 
The modeling results for WASP 0346-21 indicate the existence of the third light $l_3$ = 0.043$\pm$0.004 in the TESS data. This value 
roughly corresponds to the average contribution (4.7 \%) of the tertiary component (C) to the total system light estimated from 
the BF profiles for the red region spectra (e.g., Rucinski \& Pribulla 2008; Nelson 2022). Then, the $l_3$ source might be identified 
as WASP 0346-21 C detected in the UVES spectra. The physical properties of WASP 0346-21 AB were derived from the light and RV parameters 
as follows: masses of $M_{\rm A}$ = 1.57$\pm$0.05 M$_\odot$ and $M_{\rm B}$ = 0.19$\pm$0.01 M$_\odot$, radii of $R_{\rm A}$ = 1.78$\pm$0.04 M$_\odot$ 
and $R_{\rm B}$ = 0.31$\pm$0.02 M$_\odot$, effective temperatures of $T_{\rm eff,A}$ = 7225$\pm$42 K and $T_{\rm eff,B}$ = 10,655$\pm$146 K, 
and luminosities of $L_{\rm A}$ = 7.7$\pm$0.4 L$_\odot$ and $L_{\rm B}$ = 1.1$\pm$0.2 L$_\odot$. The low $q$ and $M_2$ combination 
of the short-period eclipsing pair is fully consistent with being an EL CVn binary (Maxted et al. 2014; Chen et al. 2017; Lee et al. 2020). 

WASP 0346-21 A and B are marked as star symbols on the H-R diagram in Figure 8. Here, the solid and dash oblique lines denote 
the instability edges of the $\delta$ Sct and $\gamma$ Dor variables, respectively. The gray and black lines are WD evolution tracks for 
metallicities $Z$ = 0.001 and 0.01 with masses between 0.183 M$_\odot$ and 0.190 M$_\odot$ (Istrate et al. 2016). As shown in this 
diagram, WASP 0346-21 A is located within an overlapping main sequence region of the two pulsation types, and its secondary companion 
of 0.185 $\pm$ 0.013 $M_\odot$ is in full agreement with the low-mass He WD model for $Z$ = 0.01 and is currently thought to be evolving 
towards higher temperatures at the constant $L$ stage. During the pre-He WD phase, the lifetime ($t$) of WASP 0346-21 B depends on 
its mass $M_{\rm WD}$, which is calculated to be $\sim$8 $\times$ 10$^{8}$ yr from the $M_{\rm WD}-t$ relation by Chen et al. (2017). 
The post-mass transfer EBs such as R CMa and EL CVn binaries have been proposed to be formed from the stable Roche-lobe overflow channel 
(Chen et al. 2017; Lee et al. 2018). This evolution channel predicts that there is a close correlation between $P_{\rm orb}$ and 
$M_{\rm WD}$ (Rappaport et al. 1995; Lin et al. 2011). WASP 0346-21 B is well-matched with the relationship between the two in a paper 
of Lin et al. (2011). This implies that WASP 0346-21 AB is a post-mass transfer binary produced through a stable mass transfer evolution 
during its red giant phase. 

To find out the population membership of WASP 0346-21, we obtained its space motion ($U, V, W$) and Galactic orbit ($J_{\rm z}$, $e$) 
using the same solar values and procedure as Lee et al. (2020). These were calculated by applying the astrometric data from the Gaia DR3 
(Gaia Collaboration 2022) and our center-of-mass $\gamma$ velocity for the eclipsing components. We plotted the kinematical data of 
our target star on the $U-V$ and $J_{\rm z}-e$ diagrams in Figure 9, together with the other double-lined EL CVn stars: 
EL CVn (Wang et al. 2020), KOI-81 (Matson et al. 2015), WASP 0131+28 (Lee et al. 2020), WASP 0247-25 (Maxted et al. 2013; Kim et al. 2021), 
WASP 0843-11 (Hong et al. 2021), and WASP 1625-04 (Lee et al. 2022b). The location of WASP 0346-21 lies within the thick-disk regions in 
the two diagrams, which are useful for the population classification of WDs (Pauli et al. 2006). The classification of WASP 0346-21 into 
the thick-disk population ($Z \approx 0.01$; Chen et al. 2021) is concurrent with the metallicity $Z$ = 0.01 obtained from the He-core WD model fit. 

Using a relation of $v_{\rm sync}$ = $2 \pi R$/$P_{\rm orb}$, we computed the synchronized rotations of the binary components to be 
$v_{\rm A,sync}$ = 97.0 $\pm$ 0.5 km s$^{-1}$ and $v_{\rm B,sync}$ = 16.8 $\pm$ 1.3 km s$^{-1}$, respectively. Our measured rotation 
$v_{\rm A}$ indicates that WASP 0346-21 A is spinning 0.82 times slower than its synchronous value $v_{\rm A,sync}$. 
The sub-synchronization mechanism is currently not well understood in the post-mass transfer EL CVn EBs (Wang et al. 2020; Kim et al. 2021). 
However, Fuller \& Felce (2023) reported that the slow rotation in close binaries could be caused by the action of a circumbinary object. 
In addition, binary evolution theories indicate that EL CVn binaries are inner close pairs of hierarchical triplets descended from 
short-period ($P_{\rm orb}<$ $\sim$3 days) main-sequence binaries. Consistent with this prediction, Lagos et al. (2020) found 
potential tertiary objects in their observations of all five EL CVn stars. These results, together with the third-light detection, 
imply that WASP 0346-21 C, discovered in the UVES spectra, is a circumbinary star that is physically associated with WASP 0346-21 AB. 
The tertiary object may have played a significant role in the evolution of the initial binary into the present closer pair through 
re-distribution of angular momentum between them (Eggleton \& Kisseleva-Eggleton 2006; Tokovinin et al. 2006). 

The echelle spectral analysis and binary modeling presented in this paper indicate that WASP 0346-21 is a triple system (AB)C, which 
comprises a post-mass transfer binary (AB) with a short period of $<$ 1 day and a distant tertiary companion (C) with a light contribution 
of 4.3 \% in the TESS band. The fundamental stellar parameters and pulsational characteristics of the eclipsing components, and their 
locations in the H-R diagram demonstrate that the low frequencies from 14 day$^{-1}$ to 39 day$^{-1}$ are mostly $\delta$ Sct pulsations 
arising from WASP 0346-21 A, and that both high frequencies of 97.996 day$^{-1}$ and 90.460 day$^{-1}$ are oscillating signals associated 
with the pre-ELM WD companion (Breger 2000; C\'orsico et al. 2019). Then, WASP 0346-21 is the first triple system to be 
spectroscopically detected and the sixth EBs to exhibit both $\delta$ Sct and pre-ELMV multiperiod pulsations, among the EL CVn stars. 
These discoveries make WASP 0346-21 a key object for studying the dynamical evolution and asteroseismology of the post-mass transfer binaries.

\acknowledgments{ }
This article is based on VLT/UVES spectra made with ESO under programme 094.D-0027(A) and photometric data collected by the TESS 
mission. We appreciate the helpful comments and suggestions of the anonymous referee. This research has made use 
of the Simbad database maintained at CDS, Strasbourg, France, and was supported by the KASI grant 2024-1-832-01 and the project 
Cooperatio - Physics of the Charles University in Prague.

\clearpage
\begin{figure}
\includegraphics[scale=1.2]{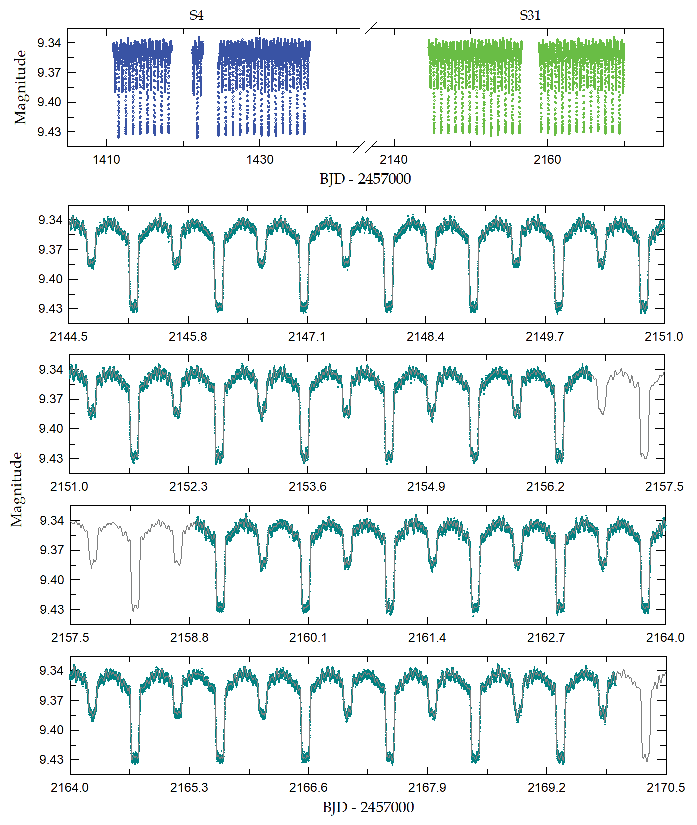}
\caption{TESS observations of WASP 0346-21 obtained during Sectors 4 and 31. The top panel represents 2-min cadence data (blue and green) 
from both sectors, and the other panels detail 20-s cadence measurements (cyan) from Sector 31. The gray lines are the sum of the two model 
curves computed from the binary star parameters of Table 3 and the 58-frequency fit of Table 5. }
\label{Fig1}
\end{figure}

\begin{figure}
\includegraphics[]{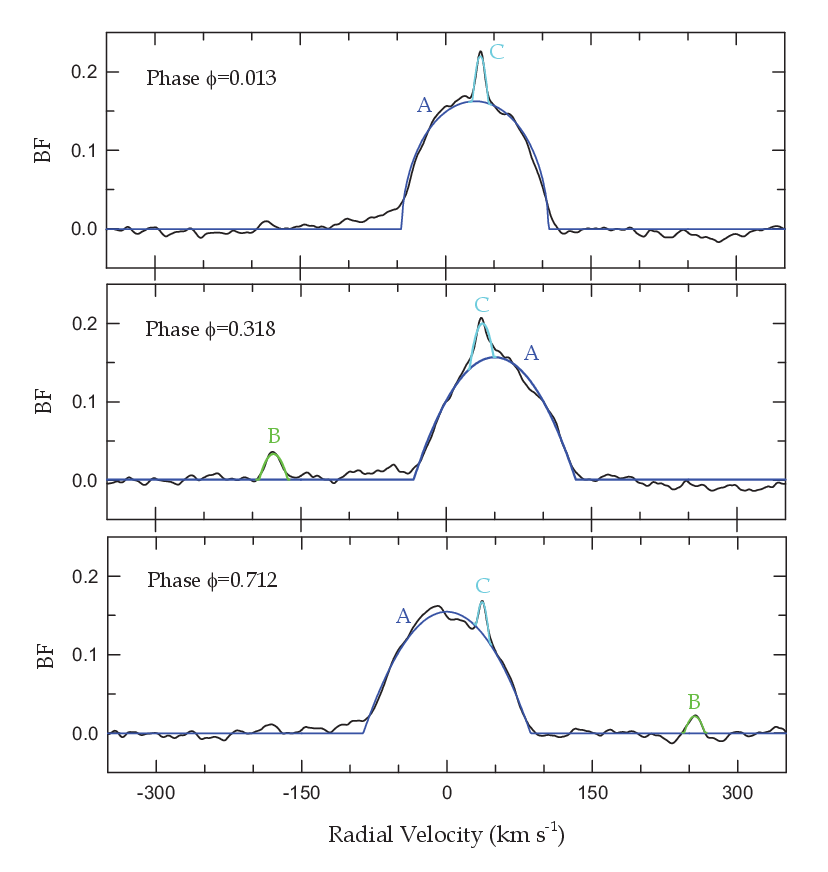}
\caption{Sample BF profiles for three orbital phases ($\phi$) obtained with the \texttt{RaveSpan} code. The solid black line indicates 
the BF profile, which is fitted with three rotational profile functions representing the primary (blue), secondary (green), and third (cyan) 
components. In the top panel, the secondary companion (B) is not visible as it is obscured by the primary component. }
\label{Fig2}
\end{figure}

\begin{figure}
\includegraphics[]{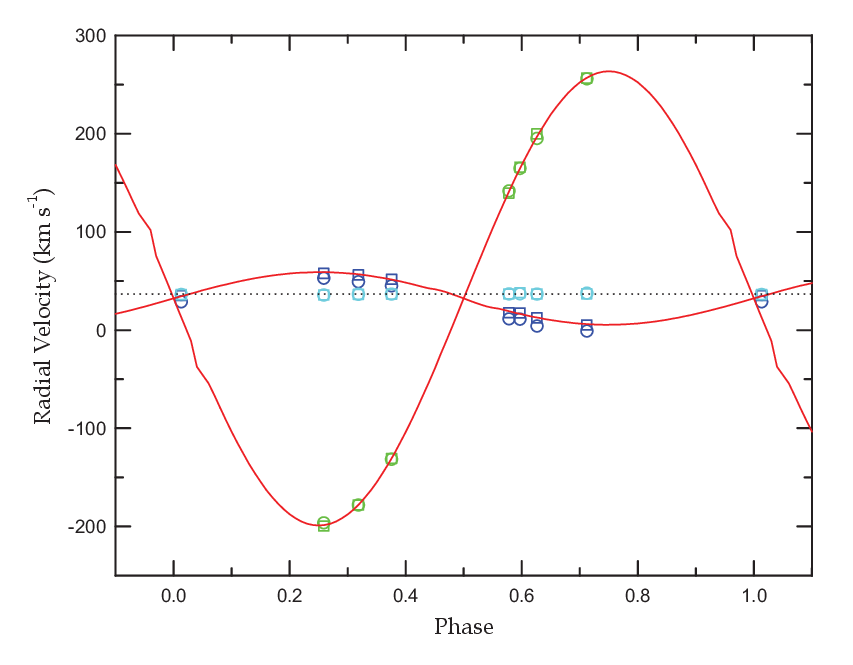}
\caption{RV curves of WASP 0346-21 with fitted models. The blue, green, and cyan colors denote the primary (A), secondary (B), and 
third (C) components, respectively. The circles and squares are the RV measurements in the blue (4440--4520 $\AA$) and red 
(6310--6390 $\AA$) spectral ranges, respectively. The solid curves represent the results of our binary model using the W-D code, and 
the dotted line denotes the average RV of $+$36.7 km s$^{-1}$ for the third component. } 
\label{Fig3}
\end{figure}

\begin{figure}
\includegraphics[scale=0.9]{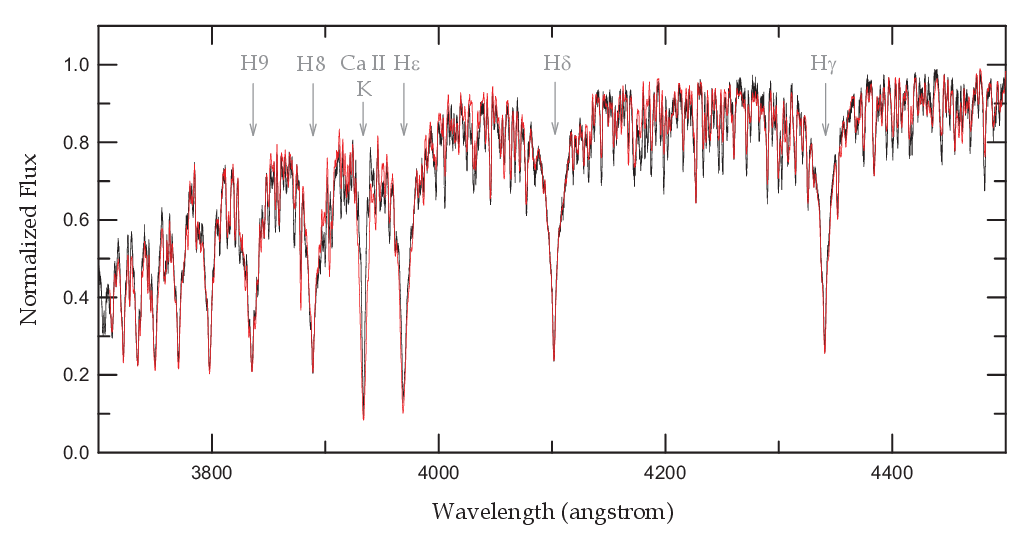}
\caption{Spectra of WASP 0346-21 at a conjunction phase of 0.013. The black and red solid lines represent the observed UVES and best-fitting 
synthetic spectra, respectively. }
\label{Fig4}
\end{figure}

\begin{figure}
\includegraphics[scale=1.3]{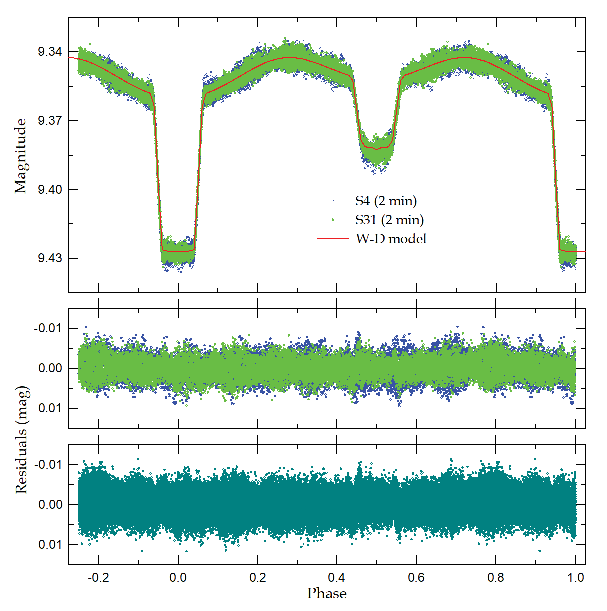}
\caption{The top panel displays the phased light curve of WASP 0346-21 with the fitted model. The blue and green circles are the 2-min 
cadence measurements from Sectors 4 and 31, respectively, and the red solid curve is computed with our W-D fit. The middle and 
bottom panels represent the corresponding residuals for 2-min and 20-s cadence data, respectively, from the synthetic curve. }
\label{Fig5}
\end{figure}

\begin{figure}
\includegraphics[scale=1.2]{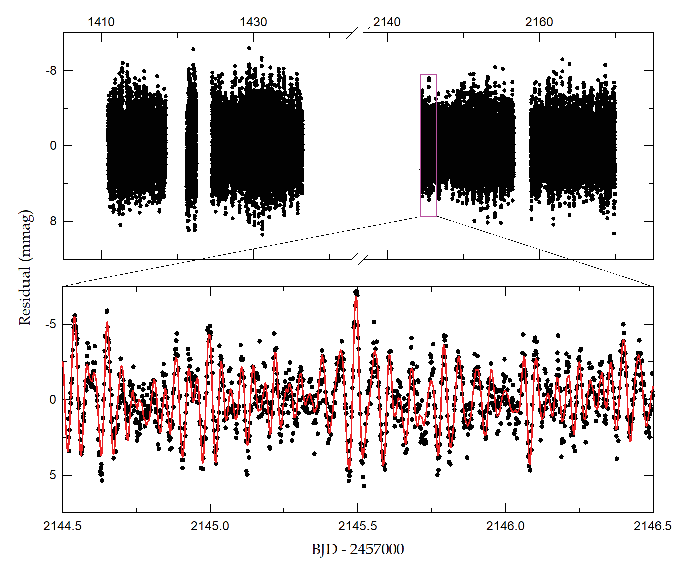}
\caption{Light curve residuals distributed in BJD after subtraction of the binarity effects from the 2-min cadence TESS observations. 
The lower panel presents a short section of the residuals marked using the inset box in the upper panel. The synthetic curve is computed 
from the 58-frequency fit to the residuals. }
\label{Fig6}
\end{figure}

\begin{figure}
\includegraphics[scale=1.2]{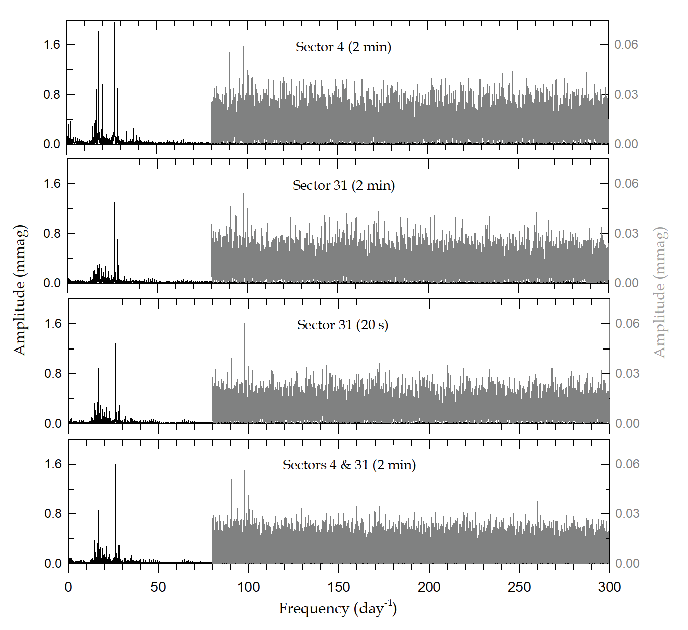}
\caption{PERIOD04 periodograms for the binary-subtracted light residuals. The first to third panels show the amplitude spectra for 
the residuals in each sector (4 and 31) and cadence (2 min and 20 s), while the bottom panel is obtained using the entire 2-min 
cadence data. In each panel, the amplitude spectrum at frequencies above 80 day$^{-1}$ is blown-up to show pre-He WD pulsations and 
is shown as a gray line. }
\label{Fig7}
\end{figure}

\begin{figure}
\includegraphics[scale=1.0]{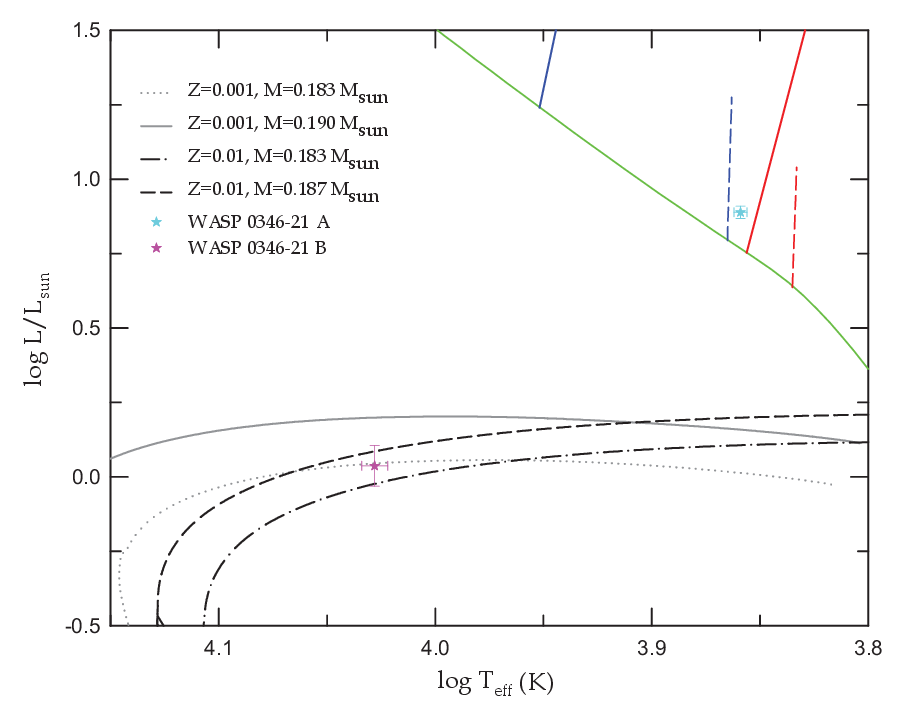}
\caption{Position in the H-R diagram of WASP 0346-21 A and B (star symbols). The green solid line is the zero-age main sequence 
(Pols et al. 1998), and the coloured oblique solid and dash lines denote the instability strips of $\delta$ Sct (Rolland et al. 2002; 
Soydugan et al. 2006) and $\gamma$ Dor stars (Warner et al. 2003; \c Cakirli 2015), respectively. The gray and black lines represent 
the evolutionary tracks of He-core WDs with different metallicities ($Z$) and masses ($M$) from the models of Istrate et al. (2016). } 
\label{Fig8}
\end{figure}

\begin{figure}
\includegraphics{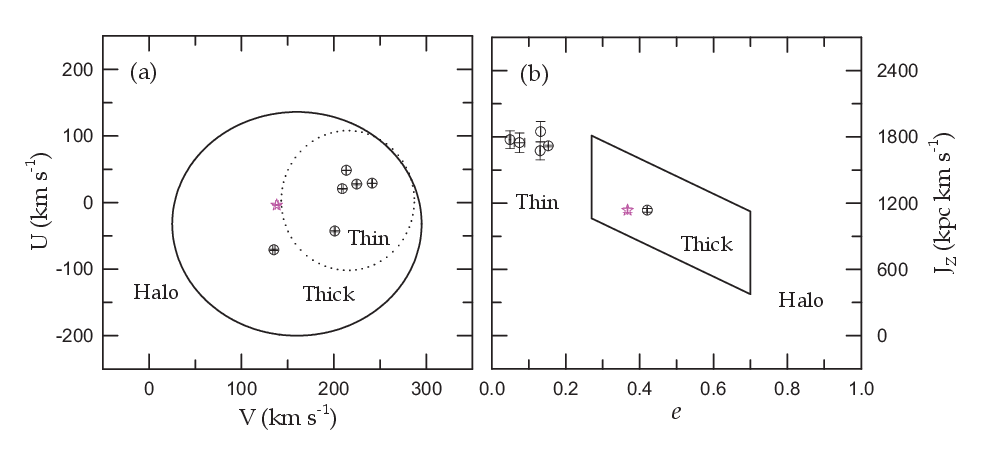}
\caption{(a) $U-V$ and (b) $J_z-e$ diagrams for WASP 0346-21 (star symbols) and the other double-lined EL CVn binaries (circles). The dot 
and solid eclipses in the left panel represent the thin and thick disk contours, respectively, and the solid box in the right panel 
indicates the thick-disk region (Pauli et al. 2006). Our target star represents a thick-disk population together with WASP 0247-25. }
\label{Fig9}
\end{figure}

\clearpage 
\begin{deluxetable}{lcccc}
\tablewidth{0pt}
\tablecaption{TESS Eclipse Timings of WASP 0346-21 }
\tablehead{
\colhead{BJD}    & \colhead{Error} & \colhead{Epoch} & \colhead{$O-C$} & \colhead{Min}  
}
\startdata
2,458,411.17012  & $\pm$0.00045	   & $-$790.5        & $+$0.00148      & II            \\
2,458,411.63412  & $\pm$0.00023	   & $-$790.0        & $+$0.00120      & I             \\
2,458,412.09702  & $\pm$0.00018	   & $-$789.5        & $-$0.00019      & II            \\
2,458,412.56058  & $\pm$0.00016	   & $-$789.0        & $-$0.00092      & I             \\
2,458,413.02630  & $\pm$0.00030	   & $-$788.5        & $+$0.00051      & II            \\
2,458,413.49122  & $\pm$0.00020	   & $-$788.0        & $+$0.00114      & I             \\
2,458,413.95577  & $\pm$0.00021	   & $-$787.5        & $+$0.00141      & II            \\
2,458,414.41820  & $\pm$0.00007	   & $-$787.0        & $-$0.00045      & I             \\
2,458,414.88220  & $\pm$0.00033	   & $-$786.5        & $-$0.00074      & II            \\
2,458,415.34739  & $\pm$0.00012	   & $-$786.0        & $+$0.00016      & I             \\
\enddata
\tablecomments{This table is available in its entirety in machine-readable form. A portion is shown here for guidance regarding its form and content.}
\end{deluxetable}

\begin{deluxetable}{lccccccc} 
\tabletypesize{\small}
\tablewidth{0pt}                    
\tablecaption{Radial velocities of WASP 0346-21$\rm ^a$ }                                                                            
\tablehead{ 
\colhead{BJD}          & \colhead{$V_{\rm A}$}   & \colhead{$\sigma_{\rm A}$} & \colhead{$V_{\rm B}$}   & \colhead{$\sigma_{\rm B}$} & \colhead{$V_{\rm C}$}   & \colhead{$\sigma_{\rm C}$} & \colhead{Region$\rm ^b$}   \\ 
\colhead{(2,456,000+)} & \colhead{(km s$^{-1}$)} & \colhead{(km s$^{-1}$)}    & \colhead{(km s$^{-1}$)} & \colhead{(km s$^{-1}$)}    & \colhead{(km s$^{-1}$)} & \colhead{(km s$^{-1}$)}    &     
}
\startdata                                                                                                                       
932.65263              & $58.22$                 & 0.23                       & $-199.66$               & 0.42                       & 35.70                   & 0.18                          & red                     \\ 
932.65280              & $52.77$                 & 0.39                       & $-196.74$               & 1.00                       & 35.86                   & 0.43                          & blue                    \\ 
932.70797              & $56.70$                 & 0.26                       & $-177.41$               & 0.48                       & 36.46                   & 0.17                          & red                     \\ 
932.70811              & $49.11$                 & 0.39                       & $-178.36$               & 0.81                       & 36.85                   & 0.53                          & blue                    \\ 
932.76111              & $52.22$                 & 0.26                       & $-130.74$               & 0.35                       & 36.71                   & 0.19                          & red                     \\ 
932.76121              & $44.95$                 & 0.38                       & $-131.87$               & 0.93                       & 36.71                   & 0.43                          & blue                    \\ 
936.66353              & $17.61$                 & 0.25                       & $ 139.44$               & 0.47                       & 36.84                   & 0.18                          & red                     \\ 
936.66363              & $11.26$                 & 0.40                       & $ 141.39$               & 0.83                       & 36.77                   & 0.44                          & blue                    \\ 
936.68108              & $16.99$                 & 0.29                       & $ 166.09$               & 0.75                       & 38.13                   & 0.24                          & red                     \\ 
936.68119              & $11.15$                 & 0.40                       & $ 164.74$               & 0.99                       & 37.03                   & 0.41                          & blue                    \\ 
936.70825              & $12.50$                 & 0.28                       & $ 199.35$               & 0.58                       & 36.85                   & 0.23                          & red                     \\ 
936.70838              & $ 3.83$                 & 0.44                       & $ 195.49$               & 1.02                       & 36.71                   & 0.37                          & blue                    \\ 
937.71650              & $ 5.16$                 & 0.30                       & $ 257.21$               & 0.41                       & 36.95                   & 0.19                          & red                     \\ 
937.71668              & $-0.61$                 & 0.40                       & $ 255.81$               & 0.99                       & 37.67                   & 0.43                          & blue                    \\ 
939.85328              & $35.78$                 & 0.22                       & $ \dots $               & \dots                      & 36.33                   & 0.16                          & red                     \\ 
939.85338              & $28.38$                 & 0.35                       & $ \dots $               & \dots                      & 36.24                   & 0.33                          & blue                    \\ 
\enddata 
\tablenotetext{a}{$V_{\rm A}$, $V_{\rm B}$, and $V_{\rm C}$ represent the measured RVs of the primary, secondary, and third stars, respectively, and $\sigma_{\rm A}$, $\sigma_{\rm B}$, and $\sigma_{\rm C}$ are their uncertainties. }
\tablenotetext{b}{Blue and red indicate the spectral regions of 4440--4520 $\AA$ and 6310--6390 $\AA$, respectively. }
\end{deluxetable}

\begin{deluxetable}{lcc}
\tablewidth{0pt} 
\tablecaption{Light and RV Parameters of WASP 0346-21 }
\tablehead{
\colhead{Parameter}               & \colhead{Primary}  & \colhead{Secondary}                                                  
}                                                                                                                                     
\startdata                                                                                                                            
$T_0$ (BJD)                       & \multicolumn{2}{c}{2,459,145.207562$\pm$0.000054}  \\
$P_{\rm orb}$ (day)               & \multicolumn{2}{c}{0.92857550$\pm$0.00000010}      \\
$i$ (deg)                         & \multicolumn{2}{c}{79.01$\pm$0.54}                 \\
$T$ (K)                           & 7225$\pm$42        & 10,655$\pm$146                \\
$\Omega$                          & 2.865$\pm$0.043    & 3.288$\pm$0.067               \\
$\Omega_{\rm in}$$\rm ^a$         & \multicolumn{2}{c}{2.012}                          \\
$F$                               & 0.819$\pm$0.071    & 1.0                           \\
$X$, $Y$                          & 0.642, 0.259       & 0.705, 0.067                  \\
$x$, $y$                          & 0.526, 0.290       & 0.398, 0.196                  \\
$l/(l_1+l_2+l_3)$                 & 0.8993$\pm$0.0028  & 0.0582                        \\
$l_3$$\rm ^b$                     & \multicolumn{2}{c}{0.0425$\pm$0.0037}              \\
$r$ (pole)                        & 0.3630$\pm$0.0060  & 0.0635$\pm$0.0046             \\
$r$ (point)                       & 0.3750$\pm$0.0069  & 0.0638$\pm$0.0047             \\
$r$ (side)                        & 0.3701$\pm$0.0065  & 0.0636$\pm$0.0046             \\
$r$ (back)                        & 0.3728$\pm$0.0067  & 0.0638$\pm$0.0047             \\
$r$ (volume)$\rm ^c$              & 0.3687$\pm$0.0065  & 0.0637$\pm$0.0047             \\ 
\multicolumn{3}{l}{Spectroscopic orbits:}                                              \\
$T_0$ (BJD)                       & \multicolumn{2}{c}{2,456,938.9125$\pm$0.0017}      \\
$a$ (R$_\odot$)                   & \multicolumn{2}{c}{4.831$\pm$0.068}                \\
$\gamma$ (km s$^{-1}$)            & \multicolumn{2}{c}{32.26$\pm$0.67}                 \\
$K_1$ (km s$^{-1}$)               & \multicolumn{2}{c}{27.2$\pm$1.7}                   \\
$K_2$ (km s$^{-1}$)               & \multicolumn{2}{c}{231.3$\pm$3.2}                  \\
$q$                               & \multicolumn{2}{c}{0.1175$\pm$0.0075}              \\
\enddata
\tablenotetext{a}{Potential for the inner critical Roche surface.}
\tablenotetext{b}{Value at 0.25 orbital phase. }
\tablenotetext{c}{Mean volume radius.}
\end{deluxetable}

\begin{deluxetable}{lcc}
\tablewidth{0pt} 
\tablecaption{Absolute Parameters of WASP 0346-21 }
\tablehead{
\colhead{Parameter}           & \colhead{Primary}   & \colhead{Secondary}                                                  
}                                                                                                                                     
\startdata                                                                                                                            
$M$ ($M_\odot$)               & 1.572$\pm$0.049     & 0.185$\pm$0.013             \\
$R$ ($R_\odot$)               & 1.780$\pm$0.040     & 0.308$\pm$0.023             \\
$\log$ $g$ (cgs)              & 4.134$\pm$0.024     & 4.729$\pm$0.072             \\
$\rho$ ($\rho_\odot$)         & 0.279$\pm$0.021     & 6.4$\pm$1.5                 \\
$v_{\rm sync}$ (km s$^{-1}$)  & 97.0$\pm$2.2        & 16.8$\pm$1.3                \\
$v$$\sin$$i$ (km s$^{-1}$)    & 78$\pm$5            & \,                          \\
$T_{\rm eff}$ (K)             & 7225$\pm$42         & 10,655$\pm$146              \\
$L$ ($L_\odot$)               & 7.74$\pm$0.39       & 1.09$\pm$0.17               \\
$M_{\rm bol}$ (mag)           & 2.50$\pm$0.05       & 4.63$\pm$0.17               \\
BC (mag)                      & 0.03$\pm$0.01       & $-$0.39$\pm$0.03            \\
$M_{\rm V}$ (mag)             & 2.47$\pm$0.05       & 5.02$\pm$0.18               \\
Distance (pc)                 & \multicolumn{2}{c}{269$\pm$8}                     \\
\enddata
\end{deluxetable}

\begin{deluxetable}{lrccrc}
\tabletypesize{\small}
\tablewidth{0pt}
\tablecaption{Results of the multiple frequency analysis for WASP 0346-21$\rm ^{a,b}$ }
\tablehead{
             & \colhead{Frequency $f_i$} & \colhead{Amplitude $A_i$} & \colhead{Phase $\phi_i$} & \colhead{SNR$\rm ^c$}  & \colhead{Remark}        \\
             & \colhead{(day$^{-1}$)} & \colhead{(mmag)}    & \colhead{(rad)} &                 &             
} 
\startdata 
$f_{1}$      &  26.34818$\pm$0.00001  & 1.619$\pm$0.032     & 4.75$\pm$0.06   & 85.94           &                          \\
$f_{2}$      &  17.68348$\pm$0.00001  & 1.559$\pm$0.037     & 4.19$\pm$0.07   & 73.11           &                          \\
$f_{3}$      &  16.65456$\pm$0.00001  & 1.166$\pm$0.037     & 3.55$\pm$0.09   & 54.10           & $f_2-f_{\rm orb}$        \\
$f_{4}$      &  19.90132$\pm$0.00001  & 0.971$\pm$0.038     & 1.75$\pm$0.12   & 43.29           & $f_3+3f_{\rm orb}$       \\
$f_{5}$      &  28.09058$\pm$0.00001  & 0.612$\pm$0.028     & 1.56$\pm$0.14   & 37.11           &                          \\
$f_{6}$      &  26.24409$\pm$0.00001  & 0.652$\pm$0.033     & 2.78$\pm$0.15   & 34.07           &                          \\
$f_{7}$      &  16.57004$\pm$0.00003  & 0.329$\pm$0.037     & 1.72$\pm$0.33   & 15.26           & $f_2-f_{\rm orb}$        \\
$f_{8}$      &  15.96747$\pm$0.00002  & 0.444$\pm$0.035     & 1.81$\pm$0.23   & 21.99           &                          \\
$f_{9}$      &  14.50208$\pm$0.00002  & 0.401$\pm$0.031     & 0.66$\pm$0.22   & 22.48           & $f_3-2f_{\rm orb}$       \\
$f_{10}$     &  17.66355$\pm$0.00002  & 0.428$\pm$0.036     & 3.02$\pm$0.25   & 20.18           & $f_2$                    \\
$f_{11}$     &   1.07421$\pm$0.00003  & 0.340$\pm$0.041     & 4.61$\pm$0.35   & 14.18           & $f_{\rm orb}$            \\
$f_{12}$     &  16.60748$\pm$0.00003  & 0.362$\pm$0.037     & 3.55$\pm$0.30   & 16.86           & $f_3$                    \\
$f_{13}$     &  26.34213$\pm$0.00002  & 0.362$\pm$0.032     & 5.05$\pm$0.26   & 19.17           & $f_1$                    \\
$f_{14}$     &  16.13082$\pm$0.00003  & 0.327$\pm$0.035     & 3.88$\pm$0.32   & 15.85           &                          \\
$f_{15}$     &  28.50201$\pm$0.00002  & 0.300$\pm$0.027     & 0.24$\pm$0.26   & 18.94           &                          \\
$f_{16}$     &   2.15245$\pm$0.00003  & 0.299$\pm$0.037     & 3.94$\pm$0.37   & 13.70           & $2f_{\rm orb}$           \\
$f_{17}$     &  19.25117$\pm$0.00003  & 0.286$\pm$0.038     & 3.20$\pm$0.39   & 12.94           & $2f_2-f_{14}$            \\
$f_{18}$     &  21.27300$\pm$0.00004  & 0.272$\pm$0.037     & 2.46$\pm$0.40   & 12.62           &                          \\
$f_{19}$     &  17.62200$\pm$0.00003  & 0.278$\pm$0.036     & 3.45$\pm$0.38   & 13.10           & $f_{10}$                 \\
$f_{20}$     &  18.95194$\pm$0.00004  & 0.240$\pm$0.039     & 0.99$\pm$0.47   & 10.57           &                          \\
$f_{21}$     &  14.42441$\pm$0.00004  & 0.214$\pm$0.030     & 2.85$\pm$0.41   & 12.15           & $f_2-3f_{\rm orb}$       \\
$f_{22}$     &  22.73966$\pm$0.00005  & 0.207$\pm$0.037     & 2.89$\pm$0.52   &  9.63           &                          \\
$f_{23}$     &  15.51922$\pm$0.00004  & 0.201$\pm$0.034     & 4.66$\pm$0.49   & 10.23           & $f_2-2f_{\rm orb}$       \\
$f_{24}$     &  17.73007$\pm$0.00004  & 0.248$\pm$0.037     & 2.31$\pm$0.43   & 11.53           & $f_2$                    \\
$f_{25}$     &  37.08501$\pm$0.00004  & 0.196$\pm$0.026     & 0.54$\pm$0.39   & 12.78           &                          \\
$f_{26}$     &  14.45765$\pm$0.00004  & 0.196$\pm$0.030     & 0.89$\pm$0.45   & 11.05           & $f_2-3f_{\rm orb}$       \\
$f_{27}$     &  33.13303$\pm$0.00004  & 0.175$\pm$0.025     & 5.48$\pm$0.42   & 12.00           & $2f_7$                   \\
$f_{28}$     &  15.06981$\pm$0.00005  & 0.173$\pm$0.033     & 5.56$\pm$0.56   &  9.01           & $f_{14}-f_{\rm orb}$     \\
$f_{29}$     &  16.62178$\pm$0.00005  & 0.201$\pm$0.037     & 0.51$\pm$0.53   &  9.38           & $f_3$                    \\
$f_{30}$     &  17.58988$\pm$0.00005  & 0.188$\pm$0.036     & 2.52$\pm$0.57   &  8.86           & $f_{19}$                 \\
$f_{31}$     &  38.76764$\pm$0.00004  & 0.162$\pm$0.025     & 1.32$\pm$0.46   & 10.92           & $2f_4-f_{\rm orb}$       \\
$f_{32}$     &  22.85869$\pm$0.00006  & 0.153$\pm$0.037     & 5.93$\pm$0.71   &  7.05           &                          \\
$f_{33}$     &  16.91802$\pm$0.00007  & 0.146$\pm$0.037     & 0.67$\pm$0.74   &  6.76           & $2f_4-f_{32}$            \\
$f_{34}$     &  15.52936$\pm$0.00005  & 0.172$\pm$0.034     & 2.71$\pm$0.57   &  8.74           & $f_2-2f_{\rm orb}$       \\
$f_{35}$     &  15.90096$\pm$0.00007  & 0.134$\pm$0.034     & 3.07$\pm$0.75   &  6.70           & $f_{31}-f_{32}$          \\
$f_{36}$     &  26.76464$\pm$0.00006  & 0.130$\pm$0.032     & 2.11$\pm$0.72   &  6.92           &                          \\
$f_{37}$     &  35.15393$\pm$0.00005  & 0.125$\pm$0.026     & 2.34$\pm$0.61   &  8.21           & $2f_{30}$                \\
$f_{38}$     &  17.22793$\pm$0.00009  & 0.109$\pm$0.036     & 2.06$\pm$0.96   &  5.24           & $f_{25}-f_4$             \\
$f_{39}$     &  21.19250$\pm$0.00008  & 0.120$\pm$0.037     & 4.56$\pm$0.90   &  5.55           & $f_{31}-f_{19}$          \\
$f_{40}$     &  26.69991$\pm$0.00007  & 0.119$\pm$0.032     & 2.94$\pm$0.79   &  6.34           &                          \\
$f_{41}$     &   3.23351$\pm$0.00008  & 0.115$\pm$0.035     & 4.20$\pm$0.89   &  5.62           & $f_4-f_3$                \\
$f_{42}$     &  15.31097$\pm$0.00008  & 0.116$\pm$0.034     & 1.68$\pm$0.85   &  5.91           &                          \\
$f_{43}$     &  36.61512$\pm$0.00006  & 0.114$\pm$0.026     & 2.36$\pm$0.66   &  7.61           & $f_{20}+f_2$             \\
$f_{44}$     &  28.67571$\pm$0.00006  & 0.113$\pm$0.027     & 0.04$\pm$0.70   &  7.18           &                          \\
$f_{45}$     &  17.69393$\pm$0.00006  & 0.161$\pm$0.037     & 3.70$\pm$0.67   &  7.51           & $f_2$                    \\
$f_{46}$     &  22.16505$\pm$0.00008  & 0.115$\pm$0.036     & 2.29$\pm$0.93   &  5.40           & $f_{31}-f_3$             \\
$f_{47}$     &   1.70530$\pm$0.00010  & 0.106$\pm$0.040     & 2.47$\pm$1.11   &  4.52           & $f_5-f_1$                \\
$f_{48}$     &  24.73803$\pm$0.00009  & 0.105$\pm$0.035     & 5.75$\pm$0.98   &  5.09           &                          \\
$f_{49}$     &  28.08824$\pm$0.00002  & 0.316$\pm$0.028     & 6.09$\pm$0.26   & 19.17           & $f_5$                    \\
$f_{50}$     &  27.99998$\pm$0.00007  & 0.109$\pm$0.028     & 0.82$\pm$0.76   &  6.56           & $f_1+f_{47}$             \\
$f_{51}$     &  22.30011$\pm$0.00009  & 0.108$\pm$0.037     & 4.33$\pm$1.00   &  5.00           & $f_{18}+f_{\rm orb}$     \\
$f_{52}$     &  17.15799$\pm$0.00007  & 0.135$\pm$0.036     & 2.11$\pm$0.78   &  6.40           & $f_{25}-f_4$             \\
$f_{53}$     &  17.08872$\pm$0.00009  & 0.112$\pm$0.037     & 4.06$\pm$0.96   &  5.21           & $f_8+f_{\rm orb}$        \\
$f_{54}$     &  13.36270$\pm$0.00007  & 0.100$\pm$0.026     & 4.77$\pm$0.77   &  6.48           & $f_{21}-f_{\rm orb}$     \\
$f_{55}$     &  34.46275$\pm$0.00007  & 0.096$\pm$0.026     & 1.25$\pm$0.79   &  6.37           & $2f_{38}$                \\
$f_{56}$     &  31.68340$\pm$0.00006  & 0.095$\pm$0.023     & 0.92$\pm$0.72   &  6.95           & $f_3+f_{28}$             \\
$f_{57}$     &  97.99553$\pm$0.00007  & 0.056$\pm$0.015     & 2.18$\pm$0.77   &  6.49           &                          \\
$f_{58}$     &  90.45982$\pm$0.00008  & 0.051$\pm$0.016     & 2.86$\pm$0.89   &  5.60           &                          \\
\enddata
\tablenotetext{a}{Frequencies, amplitudes, and phases are the values from the equation $Z$ = $Z_0$ + $\Sigma _{i}$ $A_i \sin$(2$\pi f_i t + \phi _i$). }
\tablenotetext{b}{Parameters' errors were obtained following Kallinger et al. (2008). }
\tablenotetext{c}{Calculated in a range of 5 day$^{-1}$ around each frequency. }
\end{deluxetable}

\end{document}